\documentclass[preprint,twocolumn,10pt,a4paper,byrevtex,prb,unsortedaddress,superscriptaddress]{revtex4}
\usepackage[utf8]{inputenc}
\usepackage[T1]{fontenc}

\usepackage{amssymb}
\usepackage{soul}
\usepackage{graphicx}
\usepackage{hyperref}
\usepackage{subfigure}
\usepackage{amsmath}
\usepackage{color}

\newcommand{\ve}{\boldsymbol}
\newcommand{\vek}{\boldsymbol{k}}
\newcommand{\Real}{{\mathrm{Re}}}

\definecolor{DarkGreen}{rgb}{0,0,0}
\newcommand{\farkhad}[1]{\textcolor{DarkGreen}{#1}}
\definecolor{Purple}{rgb}{0,0,0.0}
\newcommand{\lina}[1]{\textcolor{Purple}{#1}}
\definecolor{Orange}{rgb}{1,0.6,0}

\definecolor{Blue}{rgb}{0,0,0.9}
\newcommand{\cesar}[1]{\textcolor{DarkGreen}{#1}}
\begin{document}

\title{Superconductivity assisted change of the perpendicular magnetic anisotropy in V/MgO/Fe junctions}




\author{C\'esar Gonz\'alez-Ruano}
\affiliation{Departamento F\'isica de la Materia Condensada C-III, Instituto Nicol\'as Cabrera (INC) and  Condensed Matter Physics Institute (IFIMAC), Universidad Aut\'onoma de Madrid, Madrid 28049, Spain}

\author{Diego Caso}
\affiliation{Departamento F\'isica de la Materia Condensada C-III, Instituto Nicol\'as Cabrera (INC) and  Condensed Matter Physics Institute (IFIMAC), Universidad Aut\'onoma de Madrid, Madrid 28049, Spain}

\author{Lina G. Johnsen}
\affiliation{Center for Quantum Spintronics, Department of Physics, Norwegian University of Science and Technology, NO-7491 Trondheim, Norway}

\author{Coriolan Tiusan}
\affiliation{Department of Physics and Chemistry, Center of Superconductivity Spintronics and Surface Science C4S, Technical University of Cluj-Napoca, Cluj-Napoca, 400114, Romania}
\affiliation{Institut Jean Lamour, Nancy Universit\`{e}, 54506 Vandoeuvre-les-Nancy Cedex, France}

\author{Michel Hehn}
\affiliation{Institut Jean Lamour, Nancy Universit\`{e}, 54506 Vandoeuvre-les-Nancy Cedex, France}

\author{Niladri Banerjee}
\affiliation{Department  of  Physics,  Loughborough  University, Epinal  Way, Loughborough, LE11 3TU, United Kingdom}

\author{Jacob Linder}
\affiliation{Center for Quantum Spintronics, Department of Physics, Norwegian University of Science and Technology, NO-7491 Trondheim, Norway}

\author{Farkhad G. Aliev}
\email[e-mail: ]{farkhad.aliev@uam.es}
\affiliation{Departamento F\'isica de la Materia Condensada C-III, Instituto Nicol\'as Cabrera (INC) and  Condensed Matter Physics Institute (IFIMAC), Universidad Aut\'onoma de Madrid, Madrid 28049, Spain}

\begin{abstract}
Controlling the perpendicular magnetic anisotropy (PMA) in thin films has received considerable attention in recent years due to its technological importance. PMA based devices usually involve heavy-metal (oxide)/ferromagnetic-metal bilayers, where, thanks to interfacial spin-orbit coupling (SOC), the in-plane (IP) stability of the magnetization is broken. Here we show that in V/MgO/Fe(001) epitaxial junctions with competing in-plane and out-of-plane (OOP) magnetic anisotropies, the SOC mediated interaction between a ferromagnet (FM) and a superconductor (SC) enhances the effective PMA below the superconducting transition. This produces a partial magnetization reorientation without any applied field for all but the largest junctions, where the IP anisotropy is more robust; for the smallest junctions there is a reduction of the field required to induce a complete OOP transition ($H_\text{OOP}$) due to the stronger competition between the IP and OOP anisotropies. Our results suggest that the degree of effective PMA could be controlled by the junction lateral size in the presence of superconductivity and an applied electric field. We also discuss how the $H_\text{OOP}$ field could be affected by the interaction between magnetic stray fields and superconducting vortices.
Our experimental findings, supported by numerical modelling of the ferromagnet-superconductor interaction, open pathways to active control of magnetic anisotropy in the emerging dissipation-free superconducting spin electronics.
\end{abstract}

\maketitle

\section{Introduction}

Control of out-of-plane (OOP) anisotropies in ultra thin ferromagnetic multilayer films have revolutionized magnetic storage and spintronics technologies by mitigating the impact of the demagnetizing energy as the bit and magnetic tunnel junction sizes diminished \cite{perpendicular,Dieny2017}. Tuning of perpendicular magnetic anisotropy (PMA) by careful selection of structure design \cite{Chuang2019,Yi2021} and size \cite{Sun2017} has been among the main challenges of spintronics.
Besides the variation of the ferromagnet thickness and interface with oxides, OOP magnetization reorientation can be achieved by a temporary reduction of the IP-OOP barrier using, for example, heat and microwave pulses \cite{Challener2009,Zhu2008} or a combination of magnetic field and low temperature \cite{Martinez2018}.

Recently, we demonstrated a fundamentally different route to magnetization reorientation through the influence of superconductivity on the IP magnetization anisotropy \cite{GonzalezRuano2020}. The key idea behind this effect is that the magnetization aligns to minimize the weakening of the superconducting condensate associated with the creation of spin triplet (ST) Cooper pairs \cite{Johnsen2019}. The spin triplet generation depends on the magnetization direction relative to the interfacial Rashba spin-orbit field. Understanding the factors influencing this superconductivity-induced change of magnetic anisotropy is crucial for designing the next generation of cryogenic memories in the emerging field of superconducting spintronics, where control over non-volatile magnetization states still remains a major challenge \cite{Banerjee2014,Baek2014,Gingrich2016,Satchell2021}.

The main underlying physical mechanisms for the transformation of ST Cooper pairs from singlet to mixed-spin and equal-spin triplet pairs are magnetic inhomogeneities \cite{Bergeret2001,Keizer2006}, two misaligned FM layers \cite{Fominov2010,Leksin2012} or SOC \cite{Bergeret2013}. Previous experiments focusing on SOC-driven generation of triplets have focused on heavy metal (Pt) layers in non-epitaxial SC/FM structures \cite{Banerjee2018,Satchell2018} and Rashba SOC in epitaxial V/MgO/Fe junctions \cite{Martinez2020,GonzalezRuano2020} where ST Cooper pairs are generated depending on the magnetization orientation relative to the Rashba field.

Theoretically, it has been shown \cite{Johnsen2019} that a superconductor coupled to a ferromagnet by SOC could stimulate the modification not only of the IP \cite{GonzalezRuano2020}, but also of the OOP magnetic anisotropy below the superconducting critical temperature ($T_C$). Due to the stray fields, however, ferromagnetic films are expected to have a stronger interaction with the superconductor when an OOP magnetization is present, compared to a simple IP variation \cite{Dubonos2002,Milosevic2003}. Therefore, a careful consideration of the interaction of these stray field generated by the OOP magnetization and superconducting vortices is essential to fully capture the factors influencing the effective OOP anisotropy.

Here, we investigate the superconductivity-induced OOP magnetization reorientation in epitaxial Fe(001) films with competing IP and OOP anisotropies, both at zero field and in the presence of out-of-plane applied magnetic fields. The V/MgO/Fe(001) junctions are ideal candidates to verify the predicted modification of the effective perpendicular anisotropy in the superconducting state for several reasons \cite{Johnsen2019}. Firstly, the Fe(001) has the required \cite{Johnsen2019} cubic symmetry; secondly, previous studies show that the normal state IP-OOP reorientation transition takes place at a well-defined critical field \cite{Martinez2018}; thirdly, the system has Rashba type SOC, which is responsible for the PMA in MgO/Fe \cite{Yang2011}; fourthly, the relative contribution of the IP and OOP magnetization anisotropies can be tuned by changing the junction lateral size, and SOC can be varied by applying an external electric field; finally, the change in magnetization can be determined with high precision by studying the transport characteristics using a second magnetically hard Fe/Co layer which is magnetostatically decoupled from the soft Fe layer\cite{Martinez2018}.

For the smallest junctions, where the IP and OOP anisotropies strongly compete, we remarkably observe the full superconductivity-induced IP-OOP magnetization reorientation predicted in \cite{Johnsen2019}. This results in (i) a decreasing of the required field to induce the full IP-OOP transition below $T_C$, which is not observed in bigger junctions; and (ii) a spontaneous increasing of the misalignment angle between the two FM layers below $T_C$ in the absence of applied field, which is consistently observed in all but the largest junctions. These differences in the observed behaviour depending on the junctions dimensions are most likely due to the IP anisotropy becoming more dominant with increasing lateral size. We discard the magnetostatic interaction between supeconducting vortices and the FM layers as the main cause of the observed effects.

\section{Results}

\begin{figure}[tbp]
\begin{center}
\includegraphics[width=\linewidth]{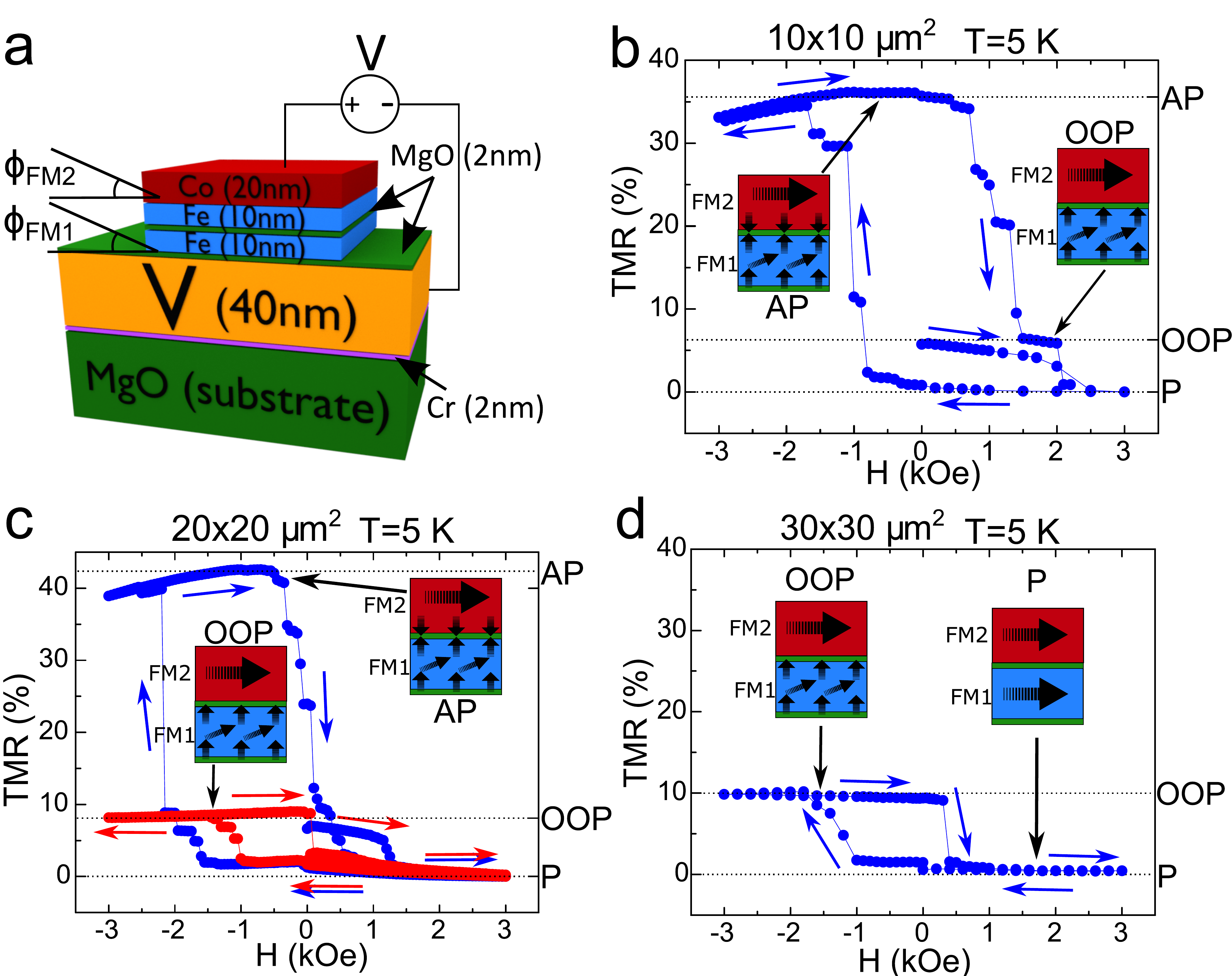}
\caption{(\textbf{a}) Sketch of the junctions under study where Fe(10~nm) (FM1) is the soft ferromagnet undergoing spin reorientation transitions, while Fe(10~nm)Co(20~nm) (FM2) is the hard (sensing) layer. $\phi_\text{FM1}$ and $\phi_\text{FM2}$ are the OOP angles of each FM layer (i.e. the angle with respect to the plane of the layers). Since the FM2 layer is normally fixed to act as a sensor, $\phi_\text{FM2}$ is assumed to be very close to 0 unless otherwise stated. (\textbf{b}), (\textbf{c}) and (\textbf{d}) show TMR experiments where the field is applied in the OOP direction in $10\times10$, $20\times20$ and $30\times30$~$\mu\text{m}^2$ junctions respectively, showing the field-induced transition into the nonvolatile OOP state. The right vertical axes indicate the parallel (P), antiparallel (AP) and OOP states for each junction, marked with dotted lines. The inset sketches depict the proposed configuration of the two FM layers in the P (only shown in  panel d), OOP and AP configurations of the spin valve stack.}
\label{Fig1}
\end{center}
\end{figure}

Figure \ref{Fig1} shows the experimental configuration and the different types of OOP transition observed above the vanadium $T_C$. Figure \ref{Fig1}~a shows the V(40~nm)/MgO(2~nm)/Fe(10~nm)/MgO(2~nm)/Fe(10~nm) /Co(20~nm) (N(SC)/FM1/FM2) junctions with a hard Fe/Co layer (FM2) sensing the magnetization alignment of the 10~nm thick Fe(001) soft layer (FM1). Details about the sample growth, characterization and the experimental set-up are explained in the Methods section. All junctions were saturated with a 3 kOe IP magnetic field (see the alignment calibration procedure in Appendix \ref{Sec:A}) before each of the OOP magnetoresistance (TMR) measurements, in order to eliminate magnetic inhomogeneities from previous OOP measurements. All except one of the studied junctions showed OOP anisotropy below 3 kOe. On the right side of the vertical axes of Figure \ref{Fig1}~b-d, we indicate the TMR values corresponding to the well established parallel (P), perpendicular out-of-plane (OOP) and antiparallel (AP) states for each sample, which are used to calibrate the angle between the two FM layers ($\Delta\phi=\phi_\text{FM1}-\phi_\text{FM2}$, where $\phi_\text{FM1}$ and $\phi_\text{FM2}$ are the angles of each FM layer with respect to the plane of the layers, as shown in Figure \ref{Fig1}~a) with the same procedure as described in Refs. \cite{Martinez2018,GonzalezRuano2020}. This indicates that the IP-OOP transition also triggers a total or partial reorientation of the sensing (hard) FM2 layer, providing a resistance close to that of an AP state. Previous OOP measurements\cite{Martinez2018} above $T_C$ made on only two $20\times20$~$\mu\text{m}^2$ junctions revealed asymmetric transitions into the perpendicular alignment of the soft FM1 layer, without any subsequent transition of the sensing layer with perpendicular fields up to 3 kOe. The present study is made with a total of 16 junctions of four different lateral sizes, where about half of them also demonstrate a transition to an AP configuration when the magnetic field is further increased after the transition to the OOP state has been completed. This AP configuration could potentially be either with the two layers oriented OOP or IP, although it seems rather unlikely that both layers reorient to an IP configuration while the applied OOP field increases. We believe that the high-field-induced transition from OOP to AP alignment or, in some cases, a nearly direct P to AP transition in N/FM1/FM2 junctions could be a consequence of the enhanced antiferromagnetic coupling reported for MgO magnetic tunnel junctions with perpendicular magnetic anisotropy (see \cite{Nistor2010}). We cannot exclude the possibility that the AP alignment could be triggered by a partial reorientation of the hard Fe/Co layer (with only the Fe part or the atomic layers closer to the Fe/MgO interface in the hard layer orienting OOP, as shown in the sketches in Figure \ref{Fig1}~b and c). However, since we measure the total resistance of the junctions, it is impossible to distinguish between these two cases from transport measurements alone. Therefore, we mainly focus on the influence of superconductivity on the transition between IP and OOP states and the partial OOP reorientation at zero magnetic field. Consequently, for the OOP field range reported here, we will assume that the $\phi_\text{FM2}$ angle of the FM2 layer with respect to the in-plane configuration is fixed and close to 0.

Figure \ref{Fig1}~c shows typical OOP TMR cycles measured in two $20\times20$~$\mu\text{m}^2$ junctions, one of them switching to an AP alignment following an OOP orientation (blue) and the other one only switches to the OOP state (red). Figure \ref{Fig1}~d shows an OOP TMR for a $30\times30$~$\mu\text{m}^2$ junction where the OOP alignment of the FM1 and FM2 electrodes remains stable up to 3 kOe. Note that all junctions showed remanent OOP alignment of the soft Fe(001) electrode once the perpendicular magnetic field is removed (Figure \ref{Fig1}). This indicates the relatively small number of interfacial defects present in our junctions, as supported by numerical simulations analyzing the OOP configuration robustness as a function of the density of interfacial defects by studying the \emph{inverse} OOP to IP transition, which are discussed in Appendix \ref{Sec:D}. 

\cesar{The symmetry broken spin reorientation observed in the OOP TMR experiments shown in 1 b-d,  has been previously explained in Ref.\cite{Martinez2018} by the difference in the dislocation density present at the top and bottom surfaces of the soft Fe(001) layer due to the growth process. This differently affects the top and bottom surface anisotropies, which leads to different intensities at each interface resulting in the magnetization being more easily reoriented into the OOP configuration for one field direction than the other. This asymmetric field behaviour might seem at odds with the Stoner-Rashba model developed in Ref.\cite{Barnes2014}. This model suggests that a net Rashba field related to the asymmetric top and bottom interfaces of a ferromagnetic film leads to a pseudo-dipolar contribution to the anisotropy which would mainly favor an in-plane magnetization, and to an uniaxial-like anisotropy favouring the perpendicular magnetization configuration. Correspondingly, the hysteresis curve of a single magnetic (here Fe) layer is expected to be an even function with respect to the external magnetic field. However, we note that the model does not fully account for the complexities discussed below that could lead to the asymmetric hysteresis we observe in our multilayer structures.} 

\cesar{The fact that the hysteresis curve is not an even function of the external magnetic field is simply related to the fact that the model is developed for a single ferromagnetic layer, while in our complex heterostructure we do not reverse the Fe/Co interface magnetization. This is not unreasonable considering a large interface anisotropy. A full magnetization reversal including interfacial magnetization would only result in an asymmetric hysteretic response. Secondly, stray field plays a relevant role in our structures and, importantly, the stray field seen by both interfaces is not similar. The bottom interface experiences the stray field of the Fe/Co top bilayer, while the top interface sees the contribution from the bottom Fe layer. In a macrospin model, increasing the stray field would decrease the perpendicular anisotropy. To fully understand the complexities of the asymmetric magnetization response, future studies, such as direct OOP magnetization measurements on the MgO/Fe/MgO structures in the absence of the sensing Fe/Co and V/MgO layers could be performed.}

It is worth mentioning a distinct feature of our junctions, having a strongly preferred IP magnetization at room temperature \cite{Martinez2018}, with the OOP configuration of the soft 10 nm Fe layer only becoming non-volatile below 80 K. In the temperature range in which this study takes place (0.3 to 7 K), the magnetic field required to induce an OOP transition in the soft layer does not typically exceed 2 kOe. These relatively low values (with respect to continuous 10 nm thick Fe films) could be explained by the combined influence of a few factors. Firstly, the variation of the relation between the IP and OOP anisotropy energies could vary with temperature, possibly favouring the OOP configuration at low temperatures \cite{Wang2020}. Secondly, interfacial strain has also demonstrated the potential to induce changes in the perpendicular anisotropy in thin ferromagnetic films \cite{Wolloch2021}. Thirdly, the IP saturation in this study was carried out with a field of 3 kOe. This value was considered sufficiently high since the resistance values were stable above 1 kOe, but it could be insufficient to induce a perfect IP alignment at low temperatures. This factor could be more relevant for the smallest junctions where edge magnetic charges would have a relatively higher influence on the measured OOP switching field, qualitatively explaining the dependence of this field with the junctions lateral size, as supported by numerical simulations (see Appendix \ref{Sec:E}, figure S\ref{Fig5SM}). Finally, as mentioned before, as long as we measure the total resistance of the junction, we can't exclude that the OOP reorientation might take place preferently in the atomic layers closer to the Fe/MgO interface (where it would be easier to reorient the magnetic moments due to the surface anisotropy). Thus, the surface OOP state (with a thickness of a few nm, close to that of the Fe magnetic exchange length \cite{Abo2013}) might be realized with the aid of interface anisotropy at the Fe/MgO interface and an external OOP magnetic field. This is shown in the sketches of the spin valve configuration in Figure \ref{Fig1}~b-d.

\subsection{Superconductivity induced change of the out-of-plane anisotropy field}

\begin{figure}[tbp]
\begin{center}
\includegraphics[width=\linewidth]{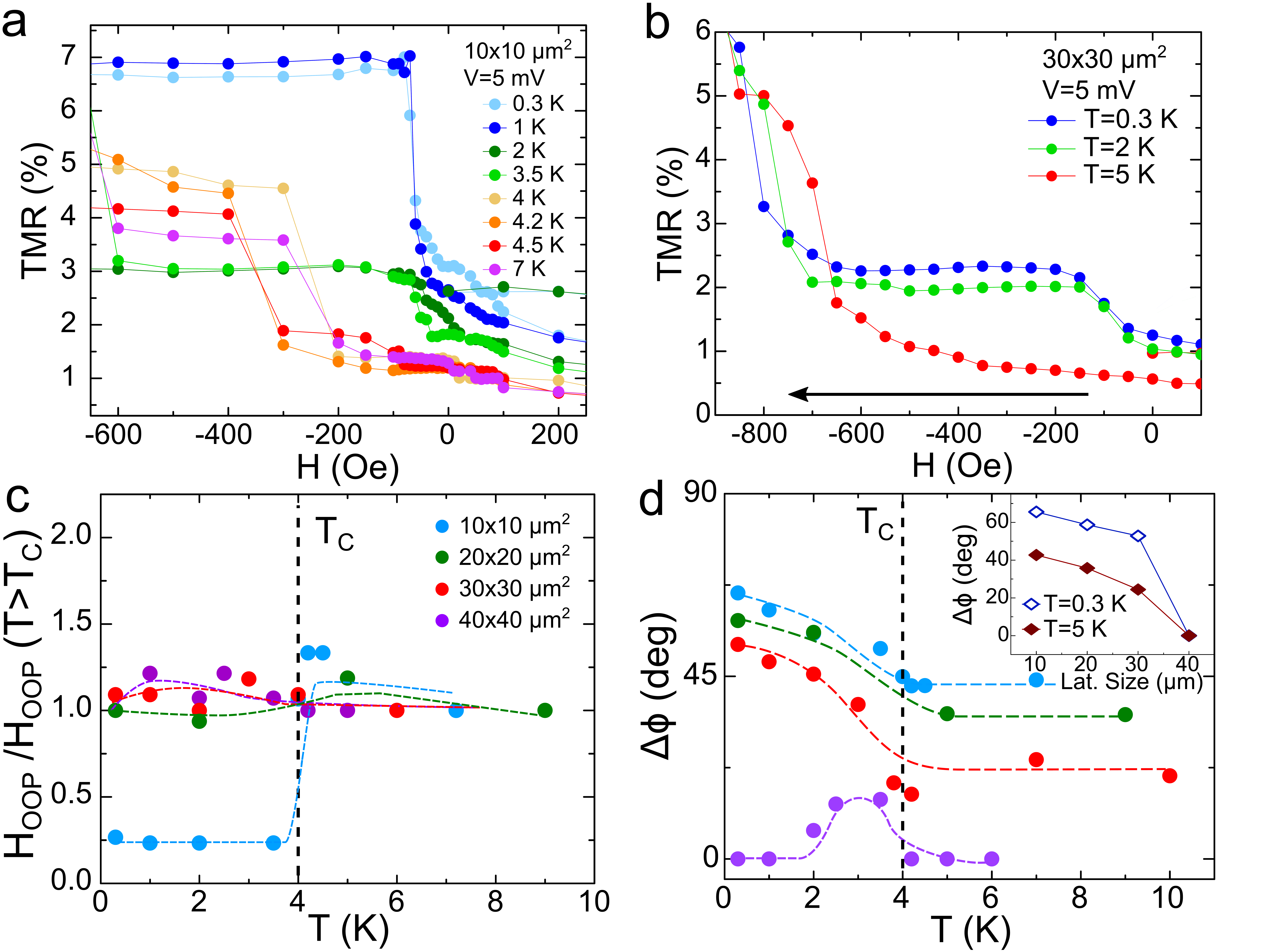}
\caption{(\textbf{a})~Field induced OOP magnetization transition in a $10\times10$~$\mu\text{m}^2$ SC/FM1/FM2 junction at different temperatures from above to below $T_C$. A strong reduction of $H_\text{OOP}$ takes place below $T_C$. (\textbf{b})~Shows a similar experiment in a $30\times30$~$\mu\text{m}^2$ junction. In this case, some increase{sting} in the low field TMR is observed, but not enough to be attributed to a complete OOP reorientation. (\textbf{c})~Temperature dependence of the normalized  $H_{\text{OOP}}$ anisotropy field for junctions with four different lateral sizes. (\textbf{d})~represents the temperature dependence of the misalignment angle between the two FM layers ($\Delta\phi=\phi_\text{FM1}-\phi_\text{FM2}$, calibrated following the procedure outlined in \cite{GonzalezRuano2020}) at zero field for the four different sized samples, using the same color legend as in (\textbf{c}). The inset shows a comparison of the zero field $\Delta\phi$ angle at $T=5$ K (above $T_C$) and at $T=0.3$ K (well below $T_C$) as a function of the samples' lateral size. The gradual decrease of the zero-field angle above $T_C$ with increasing lateral size points towards a small equilibrium initial angle already existing in the normal state, which we attribute to competing OOP and IP anisotropies. When superconductivity develops below $T_C$,  an additional magnetization reorientation is observed in all except the bigger samples. The colored dashed lines are guides for the eyes, while the vertical, black, dashed lines indicate the critical temperature.}
\label{Fig2}
\end{center}
\end{figure}

Figure \ref{Fig2}~a shows the temperature dependence of OOP TMRs in a $10\times10$~$\mu\text{m}^2$ junction, in a field range where the field-induced magnetization reorientation of the FM1 layer (measured at 5 mV) takes place. A decrease of the characteristic $H_{\text{OOP}}$ field (defined as the applied magnetic field providing a \emph{complete} OOP reorientation of the FM1 layer) just below $T_C$ can be observed upon lowering the temperature, as represented in Figure \ref{Fig2}~c. We note that the SC-induced full IP-OOP transitions have been clearly observed in the smallest $10\times10$~$\mu\text{m}^2$ lateral size junctions. The larger junctions showed a small low field TMR increase below $T_C$, which could be interpreted either as a partial FM1 layer reorientation or an inhomogeneous OOP alignment (Figure \ref{Fig2}~b). For the $20\times20$~$\mu\text{m}^2$ and larger junctions, the $H_{\text{OOP}}$ anisotropy field turned out to be nearly independent of temperature (Figure \ref{Fig2}~c). Interestingly, our junctions also revealed spontaneous zero field TMR emerging below $T_C$ (corresponding to a \emph{partial} magnetic reorientation of the soft FM1 layer), which is more pronounced for the smaller samples and diminishes with lateral size, abruptly disappearing for the largest junctions. This is shown in Figure \ref{Fig2}~d, where instead of the TMR, the calculated angle between the two FM layer is plotted. It is worth noting that this relative angle calculation is similar to our previous work\cite{Martinez2018,GonzalezRuano2020}, and assumes a uniform magnetization in the whole FM layer. However, the real scenario could be more complex (see Appendix \ref{Sec:B}).

\subsection{Influence of electric field on the out-of-plane reorientation}

The presence of the MgO barriers allows us to explore the possible influence of high electric fields on the magnetic-field-induced IP-OOP transitions above and below $T_C$. High electric field influences the PMA anisotropy by modifying the SOC Rashba field in magnetic tunnel junctions \cite{Barnes2014}. Our previous study \cite{Martinez2018} revealed that roughly two thirds of the voltage drop in our junctions occurs at the V/MgO/Fe barrier, resulting in a high electric field across this interface. The remaining voltage drops at the Fe/MgO/Fe interface, which is responsible for the change in the resistance providing the measured TMR depending on the relative magnetic configuration of the two FM layers.

\begin{figure}[tbp]
\begin{center}
\includegraphics[width=\linewidth]{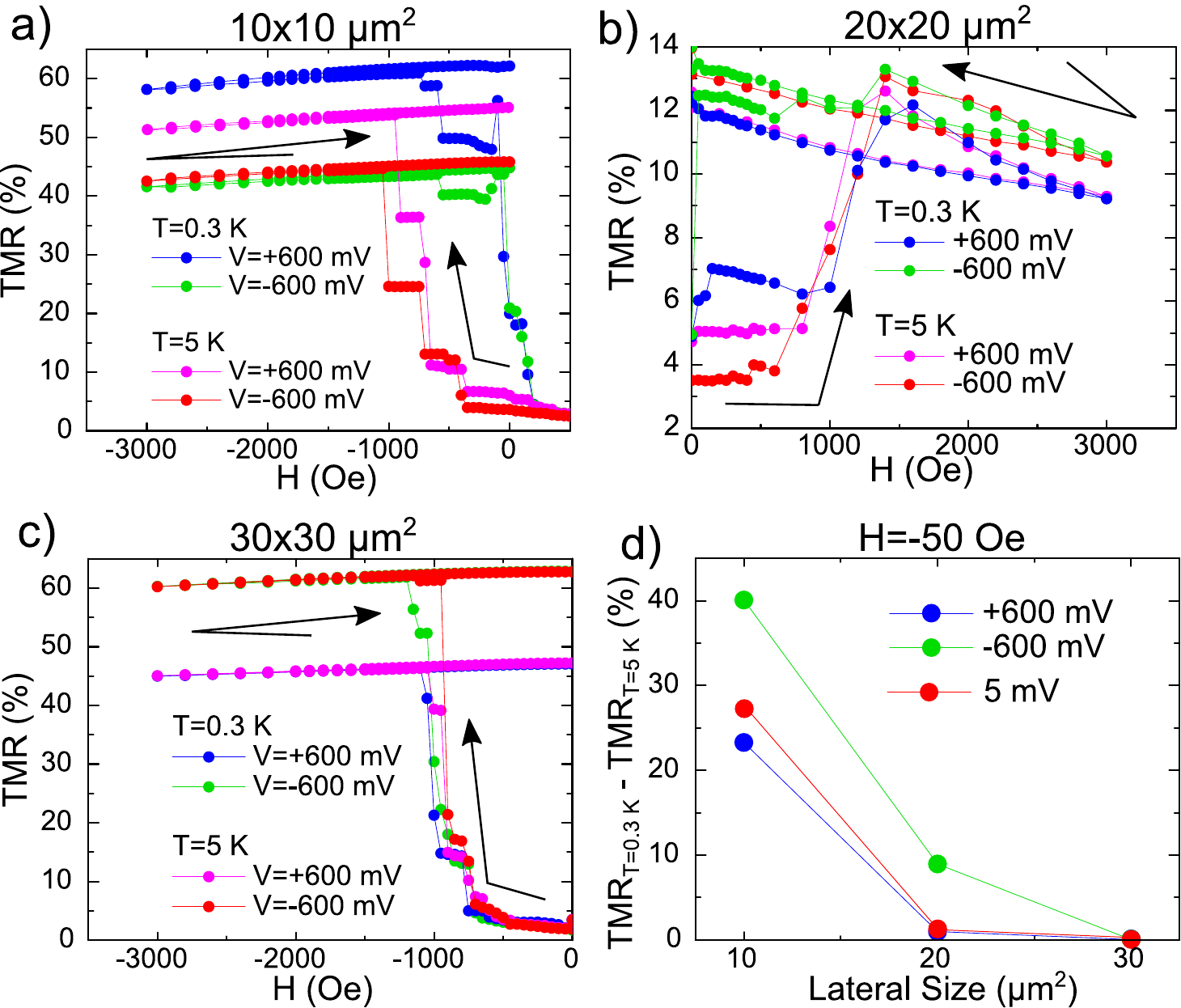}
\caption{Influence of the electric field on the magnetization reorientation transition $H_{\text{OOP}}$, above ($T=5$ K) and below $T_C$ ($T=0.3$ K). The transition is shown for an applied bias of 600~mV (electric field of about $2.5\times10^8$ V/m), with both positive and negative polarities, for samples with varying lateral sizes: (\textbf{a}) $10\times10$~$\mu\text{m}^2$, (\textbf{b}) $20\times20$~$\mu\text{m}^2$ and (\textbf{c}) $30\times30$~$\mu\text{m}^2$. (\textbf{d}) shows the difference of TMR with temperature (calculated as $\text{TMR}_{0.3\text{ K}}-\text{TMR}_{5\text{ K}}$) for both polarities and in the absence of applied electric field ($V=5$ mV) as a function of the lateral size, for an applied field of $H=-50$ Oe. The superconducting transition seems to have bigger effects on the magnetic OOP reorientation for smaller samples.}
\label{Fig3}
\end{center}
\end{figure}

We have therefore investigated the influence of high bias and its polarity on the IP-OOP transition in junctions with different lateral sizes. Figure \ref{Fig3}~a-d show that an applied bias of 600 mV (generating an electric field at the V/MgO/Fe interface exceeding $2\times10^8$ V/m) hardly affects $H_{\text{OOP}}$ above $T_C$, independently of the junctions size. Moreover, the application of a large electric field has also a negligible effect on the superconductivity-induced IP-OOP transition in the larger than $30\times30$~$\mu\text{m}^2$ junctions, with a dominant IP magnetization alignment (Figure \ref{Fig3}~c). However this changes for the smaller junctions, where IP and OOP anisotropy values are comparable leading to an entirely different behaviour. Strikingly, we observe that for $10\times10$~$\mu\text{m}^2$ and $20\times20$~$\mu\text{m}^2$ junctions, the electric field stimulates an IP-OOP transition below $T_C$ at very small values of the applied magnetic field (below 100 Oe).

Figure \ref{Fig3}~d compares the influence of an electric bias close to 600 mV with different polarities on the magnetization alignment below $T_C$ (0.3 K) with an applied magnetic field of $-50$ Oe, within the field range in which we observed a larger influence of the electric field on the IP-OOP transition for the smaller junctions. This field is about an order of magnitude below the first critical field of our Vanadium films, which was estimated to be close to 400 Oe \cite{Martinez2020}, therefore minimizing the presence of vortices in the superconducting layer. We believe that the electric field effect asymmetry could be due to the combined influence of the relatively more dominant proximity effects between the SC and FM states at the V/MgO/Fe interface in smaller junctions, and the electric-field-induced variation of the Rashba field influencing the OOP anisotropy for the non-equivalent interfaces MgO/Fe and Fe/MgO in the junctions \cite{Barnes2014}.

\section{Discussion}

\subsection{Evaluation of magnetostatic coupling between superconducting vortices and ferromagnet}
\label{Sec:3.1}

Let us start our discussion by considering different scenarios involving the possible magnetostatic coupling between the superconducting vortices and the ferromagnet \cite{Fritzsche2009}. It is tempting to consider the device edges as mainly responsible for the superconductivity-induced spin reorientation, as the edge has a more important contribution for the smallest samples, in which the minimum applied field is enough to fully reorient the magnetization. However, a few experimental facts contradict this scenario. Firstly, the superconductivity-induced additional zero field OOP angle variation is similar for $10\times10$ to $30\times30$~$\mu\text{m}^2$ junctions (see inset in figure \ref{Fig2}~d), which would not be the case if the change comes from the device's edges. The superconductivity induced spin reorientation effect abruptly diminishes for the $40\times40$~$\mu\text{m}^2$ junction only (Figure \ref{Fig2}~d). Secondly, numerical simulations show that the OOP reorientation due to magnetostatic coupling, if relevant, could potentially be triggered by the nucleation of OOP domains in the interior of the samples rather than at the edges; even if we assume the edges as the initial OOP nucleation places, the resulting vortex distribution would affect the whole FM layer (see Appendix \ref{Sec:C}). Finally, electric field stimulates the OOP transition for relatively small junctions with competing anisotropies (see Figure \ref{Fig3}) which points towards the possible role of the Rashba field.

We have seen from micromagnetic simulations and as an experimental trend that, on average, the normal state $H_{\text{OOP}}$ increases with the junctions area (Appendix \ref{Sec:E}). This is in agreement with the gradual decrease of the partial OOP magnetization reorientation with increasing lateral size seen in the normal state, just above the critical temperature (see Figure \ref{Fig2}~d).
Within the above picture, a lower $H_{\text{OOP}}$ field is required to reorient the magnetization perpendicularly in the smallest junctions, and therefore one would expect a weaker magnetostatic coupling to SC vortices.

Numerical simulations of the magnetostatic interaction of the V/MgO/Fe system during an OOP TMR experiment such as the ones shown in figure \ref{Fig1}, where a varying OOP magnetic field is applied, is a complex problem which requires self-consistent treatment of the interaction between magnetic charges and stray fields of superconducting vortices \cite{Niedzielski2019}. The Appendix \ref{Sec:C} introduces a simplified simulation scheme which evaluates this interaction in the presence of the Meissner effect. These results show that the vortex-mediated magnetostatic interaction might only explain a weak enhancement of $H_{\text{OOP}}$ in the superconducting state in the largest junctions (Figure \ref{Fig2}~c). However, we note that varying the superconducting hysteresis strength or width in the magnetostatic simulations could not explain the strong decrease of $H_{\text{OOP}}$ below $T_C$ which was experimentally observed in the smaller junctions. Moreover, a dominant magnetostatic coupling would contradict the observed influence of electric field on TMR below $T_C$ for the smallest junctions (Figure \ref{Fig2}, \ref{Fig3}).
 
\subsection{Microscopic model}

To explain the strong decrease in the OOP anisotropy field below $T_{C}$ for the smallest junctions and the superconductivity-induced zero field magnetic reorientation in all except the largest ones, as well as the influence of the SOC strength through the application of an electric field, we present a microscopic model describing the observed superconductivity-assisted OOP magnetic reorientation. In heterostructures consisting of superconducting and magnetic layers, the superconducting condensate is weakened as Cooper pairs leak into the magnetic regions \cite{Eschrig2015}. This leakage is more efficient when the spin-singlets are transformed into equal-spin triplet pairs polarized along the same axis as the magnetization. In our system, the MgO layer boosts the Rashba SOC at the SC/FM interface allowing for a generation of equal-spin triplets that depends on the orientation of the magnetization with respect to the interface \cite{Banerjee2018,Johnsen2019}.

To show how the efficiency of the triplet leakage affects the critical field for reorienting the magnetization OOP, we calculate the free energy of the system from a tight-binding Bogoliubov--de Gennes (BdG) Hamiltonian (see Appendix \ref{Sec:F} for a complete description of our method). The V/MgO/Fe structure is modelled as a cubic lattice with electron hopping between neighboring sites. We include conventional $s$-wave on-site superconducting pairing potential in the V layer, Rashba SOC in the MgO layer, and an exchange splitting between spins in the Fe layer. Although this model is valid in the ballistic limit, we expect similar results for diffusive materials since spin singlets are partially converted into odd-frequency $s$-wave triplets that are robust to impurity scattering. Moreover, the variation in the singlet population under IP to OOP reorientation of the magnetization have previously been demonstrated both experimentally and by dirty limit calculations \cite{Banerjee2018}.

The free energy determined from this model captures the contribution from the superconducting proximity effect, and also includes a normal-state contribution favoring an IP magnetization. 
In addition, we include a normal-state anisotropy $K_{\text{IP}}[1-\cos^4(\Phi_{\text{FM}1})]+K_{\text{OOP}}[1-\sin^2 (\Phi_{\text{FM}1})]$, where $\Phi_{\text{FM}1}$ ranges from 0º (corresponding to an IP magnetization of the soft ferromagnet) to 90º (OOP magnetization). In total this gives a normal-state anisotropy favoring an IP magnetization, with an additional local minimum for the OOP magnetization direction. Here we only focus on the superconductivity-assisted deepening of these OOP quasi-minima associated with the spin singlet to spin triplet conversion. The increase in $H_{\text{OOP}}$ below $T_{C}$ discussed in the previous section is not covered by this theoretical framework, as it does not take into account formation of vortices or the size of the junction. The discussion here is therefore relevant to the smaller junctions where the superconductivity-assisted decrease in $H_{\text{OOP}}$ dominates.

\begin{figure}[h]
    \centering
    \includegraphics[width=\columnwidth]{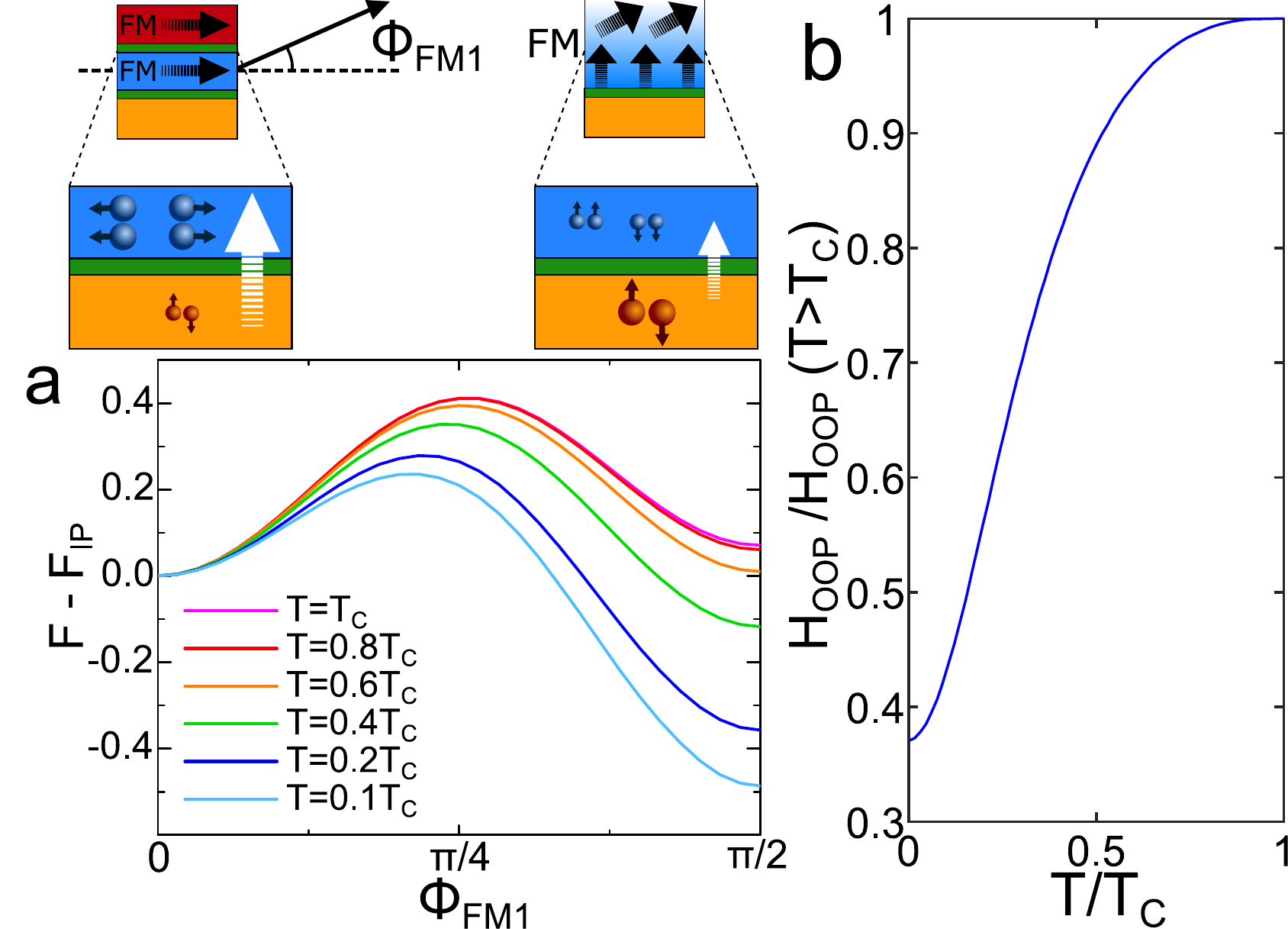}
    \caption{When the magnetization of the soft ferromagnet is rotated from a parallel to an OOP alignment with respect to the hard ferromagnet, as sketched above panel~a (In the upper right part, only the part of the soft FM layer closer to the V is depicted, not to scale, in order to show the possible magnetization configuration. In the theoretical modelling the magnetization is considered to be uniform for simplicity, altough, as mentioned before, experimentally the magnetization reorientation is more likely to happen only close to the interface. The FM2 layer is considered to be fixed with an IP orientation), the SOC assisted conversion (white arrows) of singlet Cooper pairs (orange) into equal-spin triplets (blue) is at its minimum for the OOP orientation. The superconducting condensate is therefore stronger when the magnetization is OOP, causing a decrease in the OOP free energy as the temperature is decreased below $T_C$ (panel a)). The deepening of the OOP minimum causes a decrease in $H_{\text{OOP}}$ (panel~b)). In panel~a), $K_{\text{IP}}=1.5$ and $K_{\text{OOP}}=1.6$, while in panel~b) $K_{\text{anis}}=0.8$ favoring the IP orientation. The free energy is scaled by the hopping parameter $t$. For further details about the parameters used in the BdG calculations, see Appendix \ref{Sec:F}.}
    \label{fig:BdG_theory}
\end{figure}

In Fig.~\ref{fig:BdG_theory}~a), we demonstrate how the local free energy minimum for an OOP magnetization deepens as the temperature is decreased below $T_C$. As a simple qualitative model, we calculate the external magnetic field that can be used to force the magnetization into the OOP orientation as $H_{\text{OOP}}=(K_{\text{anis}}+F_{\text{OOP}}-F_{\text{IP}})/\mu_0 \mu_{\text{tot}}$, where $\mu_{\text{tot}}$ is the total magnetic moment, $K_{\text{anis}}$ is a constant anisotropy favoring the IP orientation that includes the above mentioned parameters for the normal-state IP and OOP anisotropies, $K_\text{IP}$ and $K_\text{OOP}$, as well as an energy barrier associated with the reorientation; and $F_{\text{OOP}}$ and $F_{\text{IP}}$ are the calculated free energies in the OOP and IP states of the soft layer respectively. 
In Fig.~\ref{fig:BdG_theory}~b), we show how $H_{\text{OOP}}$ decreases below $T_C$ as observed for the $10\times10$~$\mu$m$^2$ junction in Fig.~\ref{Fig2}~c). We have thus demonstrated that the proximity effect enables a strong decrease in $H_{\text{OOP}}$ that cannot be explained by the coupling of the ferromagnet to superconducting vortices discussed in the previous section. Moreover, since this variation in $H_{\text{OOP}}$ requires that SOC is present, it also explains the dependence on the electric field observed for the smaller junctions (Fig.~\ref{Fig3}).
The fact that $H_{\text{OOP}}$ decreases over a longer temperature interval than in the experiments, rather than flattening out for low $T$, is caused by the downscaling of the lattice that is necessary in our theoretical model.  In order to scale down the superconducting coherence length so that it remains comparable to the thickness of the superconducting layer, the on-site interaction must be increased, leading to a higher $T_C$. Since the temperature interval is larger, a smaller fraction of the temperatures exist in the low-temperature limit where the free energy is temperature independent.
Keeping in mind that our measurements of $H_{\text{OOP}}$ show a dependence on the magnitude of the Rashba SOC, we can conclude that the SOC induced change in magnetic anisotropy below $T_C$ shown here must strongly contribute to the suppression in $H_{\text{OOP}}$ for the $10\times10$~$\mu$m$^2$ junctions. 

\section{Conclusions}
Our experiments point towards the superconductivity induced modification of the perpendicular magnetic anisotropy in the epitaxial Fe(001) films in the V(40~nm)/MgO(2~nm)/Fe(10~nm) system. The behaviour depends on the lateral dimensions of the junctions in the following way: First, for the smallest junctions, the magnetic field necessary for a full OOP magnetization reorientation drops by an order of magnitude in the superconducting state, while for the rest of the junctions it varies only slightly. Second, in all but the largest junctions, an increase in the OOP misalignment angle between the soft Fe(10nm) layer and the hard one is observed when the temperature is decreased below $T_C$ \emph{without any applied field}. This spontaneous reorientation is similar for $10\times10$ to $30\times30$~$\mu$m$^2$ junctions and disappears in the largest ones, suggesting that superconductivity could be affecting the competition between the IP and OOP anisotropies (which is more pronounced for the smaller junctions) rather than being the result of the reorientation taking place at the edges of the samples.
The decreasing of $H_\text{OOP}$ transition field in the superconducting state, which could also be stimulated by the application of electric field changing the Rashba SOC, is consistent with the theoretical prediction \cite{Johnsen2019} of the absolute minimum of free energy corresponding to the OOP spin direction in SC/SOC/FM hybrids with competing (IP vs OOP) anisotropies just below $T_C$. The magnetostatic interaction between vortices and magnetic inhomogeneities could explain a weak hardening of the OOP transition in the largest junctions. A detailed theoretical analysis of the mutual interplay between the inhomogeneous magnetization of the soft ferromagnet and the superconductor is, however, beyond the scope of this work. Our results open a route to active manipulation of perpendicular magnetic anisotropy in the expanding field of dissipation-free superconducting electronics involving spin \cite{Palermo2020, Shafraniuk2019, Golod2019} or spin polarized supercurrents \cite{Jeon2019}.

\section{Methods}

\subsection{Samples growth and characterization} The V(40~nm)/MgO(2~nm)/Fe(10~nm)/MgO(2~nm) /Fe(10~nm)/Co(20~nm) MTJ multilayer stacks have been grown by molecular beam epitaxy (MBE) in a chamber with a base pressure of $5\times10^{-11}$ mbar following the procedure described in Ref. \cite{Tiusan2007}. The samples were grown on (001) MgO substrates. A 10 nm thick seed of anti-diffusion MgO underlayer is grown on the substrate to trap the C from it before the deposition of the Fe (or V). The MgO insulating layer is then epitaxially grown by e-beam evaporation up to a thickness of approximately $\sim 2$ nm and the same process is then executed for the rest of the layers. Each layer is annealed at 450 ºC for 20 min for flattening. After the MBE growth, all the MTJ multilayer stacks are patterned in micrometre-sized square junctions by UV lithography and Ar ion etching, controlled step-by-step \textit{in situ} by Auger spectroscopy.

\subsection{Experimental measurement methods} The measurements are performed inside a JANIS$^{\tiny{\textregistered}}$ He$^3$ cryostat (the minimum attainable temperature is 0.3 K). The magnetic field is varied using a 3D vector magnet consisting of one solenoid (Z axis) with $H_\text{max}=3.5$ T and two Helmholtz coils (X and Y axis) with $H_\text{max}=1$ T. In our system the different magnetic states can be distinguished by looking at the resistance, so the relative orientation between two electrodes can be measured. The magnetoresistance measurements are performed by first setting the magnetic field to the desired value, then applying positive and negative current up to the desired voltage (5 mV unless otherwise stated), and averaging the absolute values of the measured voltage for the positive and negative current, obtaining a mean voltage which is used to calculate the resistance at that point. The temperature is measured and controlled with a LakeShore 340 thermometer.

\section{Acknowledgements}
We acknowledge Antonio Lara and Miguel Granda for help with simulations and Yuan Lu for help in sample preparations. The work in Madrid was supported by Spanish Ministerio de Ciencia (RTI2018-095303-B-C55) and Consejer\'ia de Educaci\'on e Investigaci\'on de la Comunidad de Madrid (NANOMAGCOST-CM Ref. P2018/NMT-4321) Grants. FGA acknowledges financial support from the Spanish Ministry of Science and Innovation, through the ``Mar\'ia de Maeztu'' Program for Units of Excellence in $R\&D$ (CEX2018-000805-M) and ``Acci\'on financiada por la Comunidad de Madrid en el marco del convenio plurianual con la Universidad Aut\'onoma de Madrid en L\'inea 3: Excelencia para el Profesorado Universitario''. D.C. has been supported by Comunidad de Madrid by contract through Consejer\'ia de Ciencia, Universidades e Investigaci\'on y Fondo Social Europeo (PEJ-2018-AI/IND-10364). NB was supported by EPSRC through the New Investigator Grant EP/S016430/1. The work in Norway was supported by the Research Council of Norway through its Centres of Excellence funding scheme grant 262633 QuSpin. C.T. acknowledges ``EMERSPIN'' grant ID PN-IIIP4-ID-PCE-2016-0143, No. UEFISCDI: 22/12.07.2017 and ``MODESKY'' grant ID PN-III-P4-ID-PCE-2020-0230 No. UEFISCDI: 4/04.01.2021. The work in Nancy was supported by CPER MatDS and the French PIA project ``Lorraine Universit\'e d'Excellence'', reference ANR-15-IDEX-04-LUE.

\appendix

\section{Correction of the OOP field misalignment}
\label{Sec:A}

\begin{figure}[ht]
\begin{center}
\includegraphics[width=\linewidth]{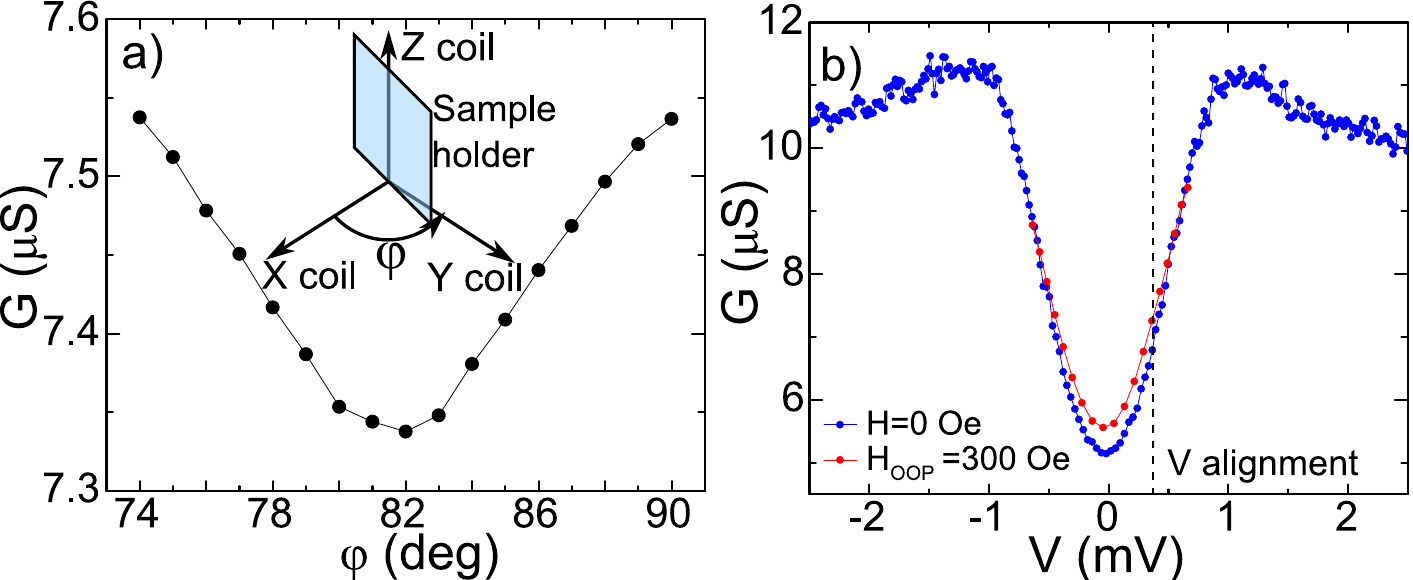}
\caption{Calibration measurements for the X axis SC coil in a $20\times20$~$\mu\text{m}^2$ junction at $T=0.3$ K. (a) shows the conductance measured at $V=0.3$ mV, inside the SC gap, for different values of the $\varphi$ angle (defined in the inset, which is a sketch of the sample holder situation with respect to the three SC coils). (b) shows two conductance curves at $H=0$ and $H_\text{OOP}=300$ Oe. The dashed line indicates the applied voltage during the calibration process. When an OOP magnetic field is present, the SC gap diminishes and the conductance increases. The misalignment angle is therefore obtained as the one which minimizes the conductance ($\varphi=82$ deg in (a)).}
\label{misalignment}
\end{center}
\end{figure}

In order to ensure that the magnetic field is perfectly aligned with respect to the FM layers, we calibrate the angle between the V/MgO/Fe plane and the magnetic field created by the vector magnet superconducting coils by performing sub-gap conductance measurements at $T=0.3$ K for different field directions around each axis. Figure \ref{misalignment} describes the calibration process in detail. The results are robust throughout the studied samples, with a misalignment of 8 $\pm1$ degrees with respect to the X-axis superconducting coil, which is accounted for in the OOP experiments. There is no observed in-plane misalignment (Y and Z axis coils).

For the highest OOP-AP transition fields measured (about 2000 Oe as shown in Figure \ref{Fig1}), the uncertainty of 1 degree in the misalignment calibration for the X coil could result in an IP component of the applied field of 35 Oe, which is about 20 times smaller than the IP coercive field of the hard FM layer \cite{GonzalezRuano2020}. Therefore, undesired IP field due to misalignment can be ruled out as an explanation for the observed AP transitions.

\section{Numerical simulations}
\label{Sec:B}

Micromagnetic simulations were carried out using the MuMax3 \cite{mumax} software \farkhad{in order to estimate (i) the possible influence of Meissner effect on the OOP transition dynamics, (ii) the role of magnetic inhomogeneities caused by defects on the volatility of the OOP magnetic state (characterized by the OOP to IP transition field, $H_\text{OOP-IP}$) and (iii) the OOP reorientation dynamics. Additional simulations of the OOP transition were performed varying the lateral size of the simulated samples, in order to contrast the results with the observed enhancement of the OOP transition field $H_\text{OOP}$ with the junctions' lateral size.} The following typical Fe magnetic parameters were implemented on the system: saturation magnetization $M_\text{S}=2.13$ T ($1.7\times 10^6$ A/m), exchange stiffness $A_\text{exch}=2.1\times 10^{-11}$ J/m, damping $\alpha=0.02$ and first order cubic anisotropy $K_{C1}=4.8\times 10^{4}$ J/m$^3$. All simulations were made with $T=0$ K. The simulation cells of the system were set at 128 for the IP components and at 16 for the OOP component. The lateral size of the simulated samples was $0.4 \times 0.4$~$\mu\text{m}^2$, with a thickness of 10 nm. The system was additionally surrounded by a vacuum box of an added 100 nm (50 nm on each side) and 3 nm on the top and bottom of the sample, leaving the discretization size of the simulation at $3.9 \times 3.9 \times 1$ nm. Higher discretization in the OOP direction has been chosen on purpose to observe the OOP effects with high accuracy in the simulations. Perpendicular magnetic anisotropy (PMA) was introduced on the top and bottom layers of the Fe as surface anisotropy, with a value of $K_{S1}=8.32\times10^{-3}$ J/m$^2$. \farkhad{Due to computational limitations on the size and detail of the simulations, the results discussed in this section should be taken as qualitative support for the experimental results, rather than quantitative estimations.}

\subsection{Influence of Meissner effect on the OOP magnetization reorientation}
\label{Sec:C}

\begin{figure}[ht]
\begin{center}
\includegraphics[width=\linewidth]{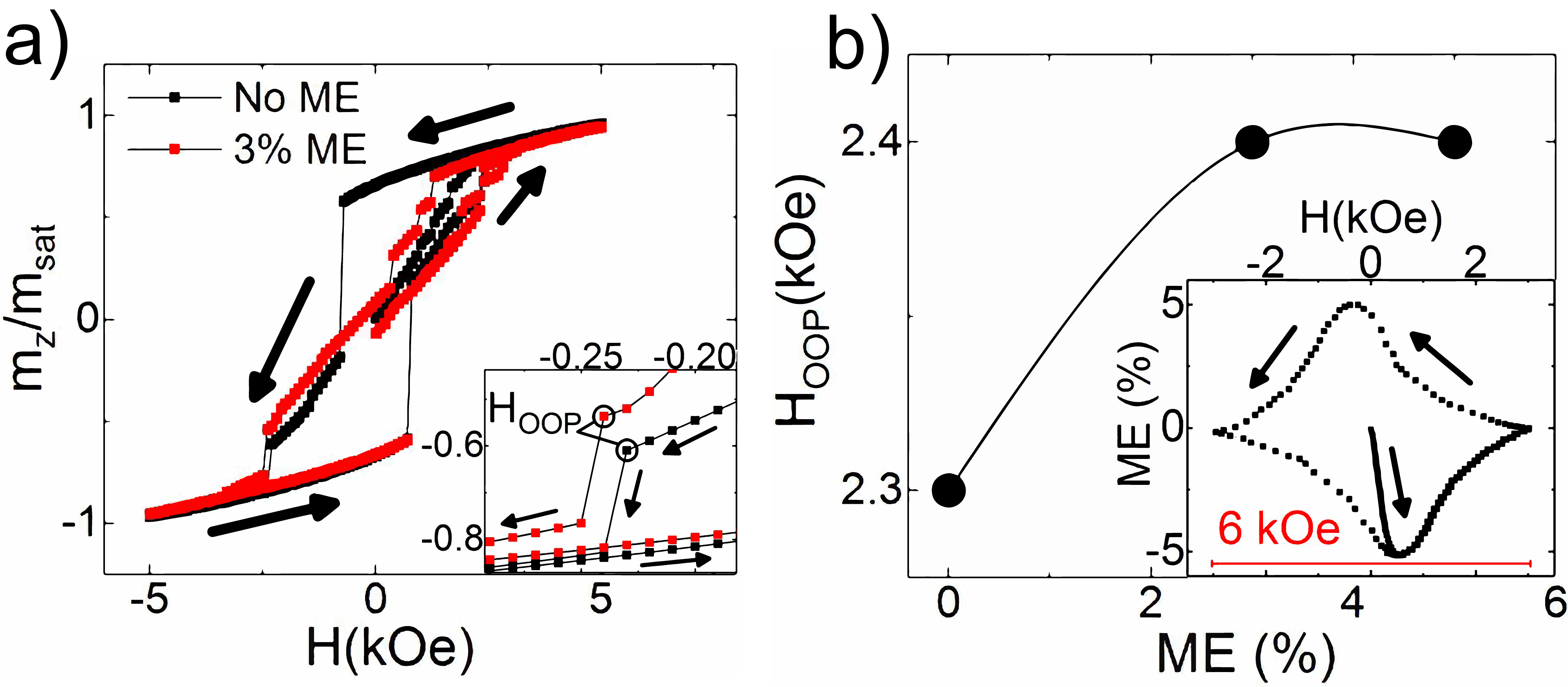}
\caption{(a) Numerical simulations of the hysteresis loops reproducing the IP to OOP transition for different strengths of ME. (b) shows the simulated OOP transition field $H_{\text{OOP}}$ as a function of the strength of the ME.}
\label{Fig2SM}
\end{center}
\end{figure}

The influence of superconductivity on the OOP transition has been studied by performing micromagnetic simulations with MuMax3 for 10 nm thick, $0.4\times0.4$~$\mu\text{m}^2$ Fe(001) films under the influence of a superconducting vanadium layer. We simulated OOP hysteresis cycles where a correction to the applied field was added based on a typical Meissner effect (ME) hysteresis cycle (obtained from \cite{Flukiger2012}, shown in Figure \ref{Fig2SM}b inset), scaled for different values of field contribution from Meissner effect and adapted to the first and second critical fields of vanadium (correspondingly $H_{c1}$ and $H_{c2}$). The contribution from superconducting vortices was taken into account by using an in-group developed program that numerically solves the time dependent Ginzburg-Landau equations in order to simulate the behaviour of type II superconductors under magnetic fields \cite{Lara2020}. The initial stray fields from an in-plane saturated FM simulation were used to generate a distribution of vortices, and then the fields generated by those vortices were calculated \cite{Chang1992} and added into the corrected hysteresis cycle. Our simplified numerical model, although limited, provides qualitative support for the mutual magnetostatic interaction between the FM and SC as the possible origin of the \farkhad{behaviour} of the $H_{\text{OOP}}$ field in the superconducting state. We also note that simulations with a contribution exceeding $7\%$ of Meissner effect resulted in the OOP state being volatile in the hysteresis cycle (i.e. the magnetization returns to an IP configuration before returning to zero field), in contradiction with the experimental observations. Consequently, we have not considered larger contributions of Meissner effect as a possible explanation for the observed behaviour of $H_{\text{OOP}}$. We also underline that a complete numerical solution is a great challenge which is outside our current capabilities, as the problem should be solved self-consistently, so the results should be understood in a qualitative way.

\begin{figure}[ht]
\begin{center}
\includegraphics[width=\linewidth]{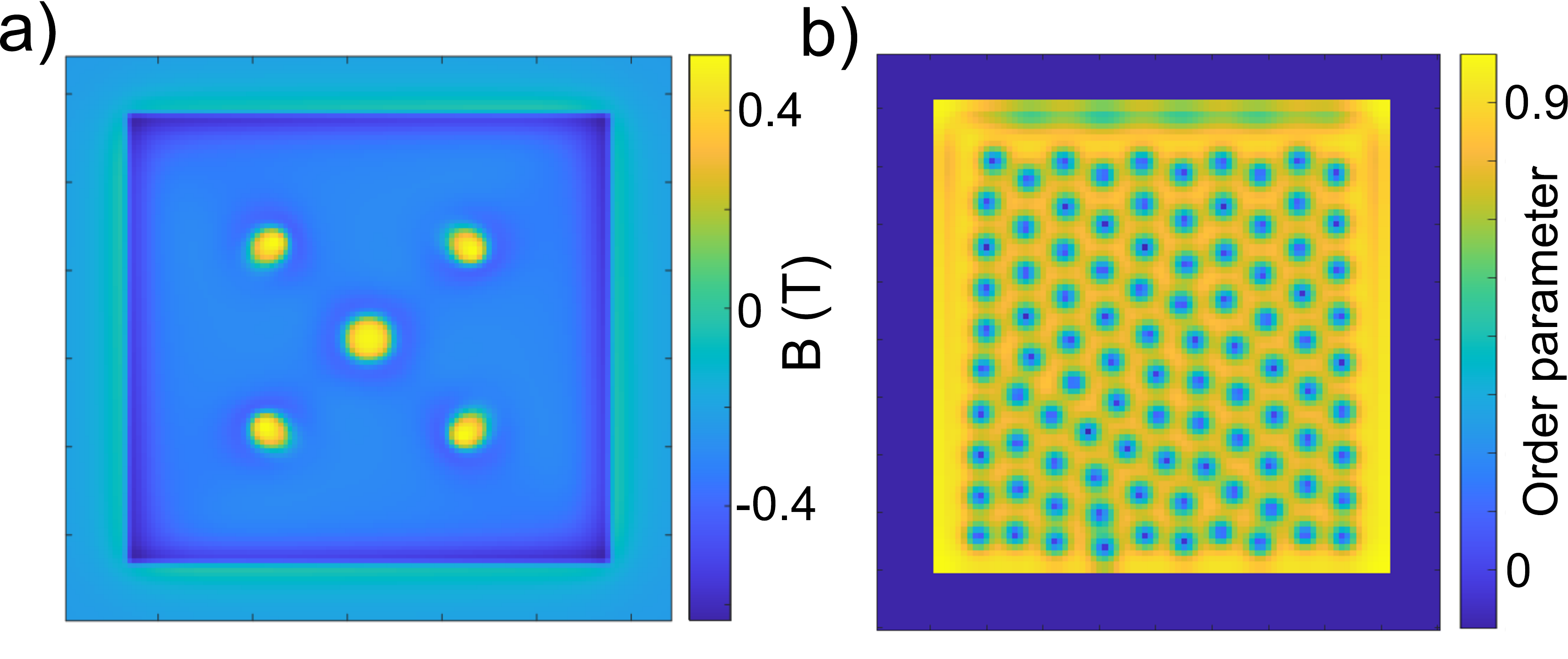}
\caption{\farkhad{(a) Snapshot of the stray fields calculated in a micromagnetic simulation of a $0.4\times0.4$ micron Fe sample behaviour under OOP applied fields just before the OOP reorientation. Five domains can be observed with an OOP magnetization resulting in high stray fields, which trigger the reorientation in the rest of the film. (b) shows the stationary state reached in a superconducting simulation on a $0.4\times0.4$ micron vanadium layer at $T=2$ K when stray fields are present at the edges of the sample. The vortices fill the interior of the film, which would result in higher OOP fields affecting the neighbouring FM layer and therefore triggering the OOP reorientation.}}
\label{StrayFields}
\end{center}
\end{figure}

\farkhad{Another concern about the OOP reorientation is the possibility of it being a trivial effect produced at the edges of the FM layers. As mentioned in the main text, there is experimental evidence pointing against this possibility. However, a more thorough study has been performed in order to fully discard this scenario. First, we simulated the field-induced OOP reorientation in $0.4\times0.4$ micron Fe films. As shown in figure \ref{StrayFields}a, the reorientation seems to be triggered by OOP oriented domains in the \emph{interior} of the film.}

\farkhad{On the other hand, we wanted to see the influence of any possible dominating edge effects on the underlying SC layer. Simulations of $0.4\times0.4$ micron V films were made using the code described in Appendix \ref{Sec:C} at $T=2$ K, with OOP magnetic fields applied at the edges of the simulated samples. As shown in figure \ref{StrayFields}b, the stationary state is reached when vortices fill the interior of the film. This would produce high OOP stray fields affecting the interior of the Fe layer, triggering a reorientation in the whole film rather than limiting it to the edges.} 

\subsection{Influence of defects on the OOP-IP magnetization reorientation}
\label{Sec:D}

\begin{figure}[tbp]
\begin{center}
\includegraphics[width=\linewidth]{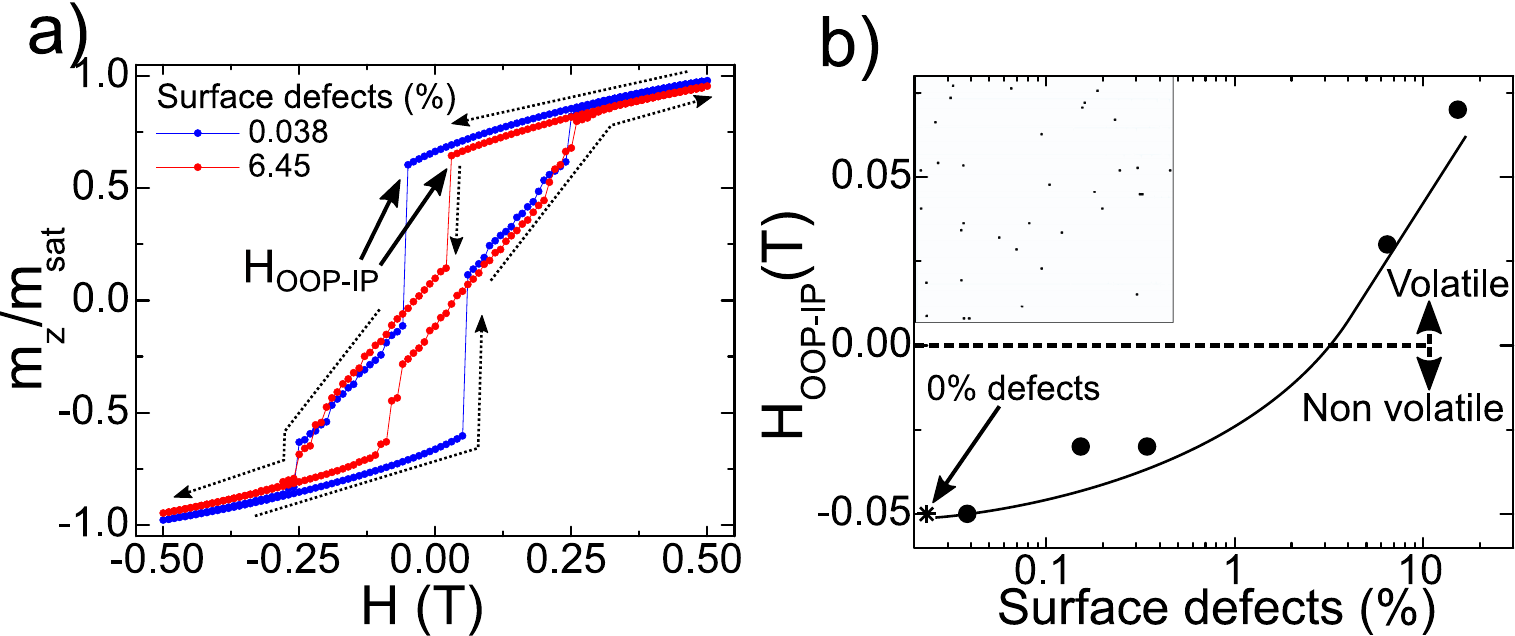}
\caption{Influence of defects on the spin reorientation transition. (a)~shows two OOP hysteresis cycles with different amounts of surface defects. The OOP-IP transition field is indicated by solid line black arrows, while the arrows with dotted lines mark the direction of the cycle. (b)~plots the transition field $H_\text{OOP-IP}$ against the $\%$ of simulated defects (in log scale). An asterisk has been manually added with the transition field for the simulation with no defects. For simulations with more than $1\%$ of defects, the OOP-IP transition becomes volatile (i.e. it happens before the field changes from positive to negative), in contradiction with the experimental results. The solid line is a guide for the eye, while the inset shows an example image of the surface defects introduced.}
\label{Fig4SM}
\end{center}
\end{figure}

 \farkhad{In order to better understand the experimentally observed non-volatility of the OOP state in the junctions,} we have simulated numerically the influence of randomly distributed surface magnetization defects within the bottom and top layers of the 10 nm thick Fe \farkhad{layer}. The defects are introduced in the simulations as spots of enhanced surface saturation magnetization ($M_S(\text{defects})=1.25\times M_S(\text{Fe})$). We have found that the introduction of a small number (about $ 10^{-3}\%$) of magnetic defects per layer does not affect the non-volatity, but only \farkhad{varies the characteristic field $H_\text{OOP-IP}$ of the transition from the OOP state to the IP alignment that takes} place after the initial magnetization saturation (Figure \ref{Fig4SM}). Above some critical defect number of about $2\times 10^{-3}\%$ defects per layer the OOP-IP transition becomes volatile. As long as we always observe experimentally non-volatility of this transition in our junctions, we can conclude that there is a relatively small number of magnetic defects present in the epitaxial MTJs under study.

\subsection{Dependence of $H_{\text{OOP}}$ with the junctions lateral size}
\label{Sec:E}

\begin{figure}[ht]
\begin{center}
\includegraphics[width=\linewidth]{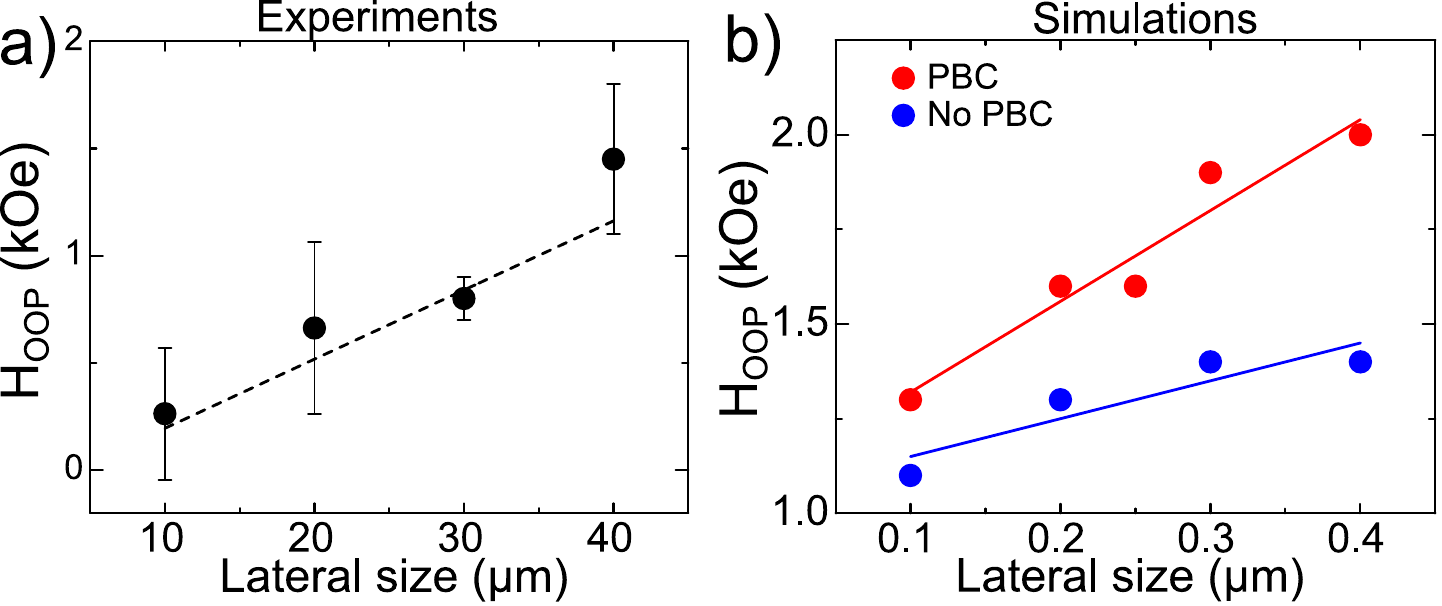}
\caption{Dependence of the OOP transition field with the lateral size of the junctions. (a) shows experimental results, with the error bars corresponding to the standard deviation of the measured samples for each size. (b) shows the transition field for micromagnetic simulations of different lateral sizes. Note that, due to computational limitations, the simulated sizes are smaller than the actual samples. However, the trend is in qualitative agreement with the experimental results. The lines are linear fittings of the experimental/simulation points, which should serve as guides for the eyes. This trend is accomplished both with or without considering periodic boundary conditions (PBC).}
\label{Fig5SM}
\end{center}
\end{figure}

The behaviour of the OOP transition field $H_{\text{OOP}}$ has been studied as a function of the samples lateral size, in a total of 14 different samples varying from $10\times10$ to $40\times40$~$\mu\text{m}^2$. We found an increasing trend of $H_{\text{OOP}}$ with the lateral size, as shown in Figure \ref{Fig5SM}a. Micromagnetic simulations of the same transition have been performed in Fe films with lateral sizes of $0.1\times0.1$ to $0.4\times0.4$~$\mu\text{m}^2$ (as the real dimensions were computationally prohibitive to simulate) with a qualitative agreement to the experimental results, as shown in Figure \ref{Fig5SM}b. \farkhad{These simulations have been made in the absence of previously discussed phenomena such as defects or Meissner effect.}

\section{Bogoliubov--de Gennes theoretical model}
 \label{Sec:F}

\lina{In order to demonstrate how the superconducting proximity effect can cause a change in the magnetic anisotropy, we consider a tight-binding Bogoliubov--de Gennes model for the V/MgO/Fe heterostructure. This model enables us to calculate the free energy of the system, so that we can study how the energy cost for reorienting the magnetization changes with temperature. We here only focus on the superconductivity-induced decrease in $H_{\text{OOP}}$ observed for $10\times10$~$\mu$m junctions. Vortex formation and the size of the lattice is not taken into account here, and instead discussed in Section~\ref{Sec:3.1} and in Appendix \ref{Sec:C}.}
For the microscopic model of the V/MgO/Fe heterostructure, we consider the Hamiltonian
\begin{align}
        \label{Hamiltonian}
        H =& -t\sum_{\left<\boldsymbol{i},\boldsymbol{j}\right>,\sigma}c_{\boldsymbol{i},\sigma}^\dagger c_{\boldsymbol{j},\sigma}
        -\sum_{\boldsymbol{i},\sigma} (\mu_{\boldsymbol{i}}-V_{\boldsymbol{i}}) c_{\boldsymbol{i},\sigma}^\dagger c_{\boldsymbol{i},\sigma}\\\nonumber 
	&-\sum_{\boldsymbol{i}} U_{\boldsymbol{i}}n_{\boldsymbol{i},\uparrow}n_{\boldsymbol{i},\downarrow}
        +\sum_{\boldsymbol{i},\alpha,\beta}c_{\boldsymbol{i},\alpha}^{\dagger}(\boldsymbol{h}_{\boldsymbol{i}}\cdot\boldsymbol{\sigma})_{\alpha,\beta} c_{\boldsymbol{i},\beta}\\\nonumber
        &-\frac{i}{2}\sum_{\left< \boldsymbol{i},\boldsymbol{j}\right> ,\alpha,\beta} \lambda_{\boldsymbol{i}} c_{\boldsymbol{i},\alpha}^{\dagger} \ve{n} \cdot (\boldsymbol{\sigma}\times\boldsymbol{d}_{\boldsymbol{i},\boldsymbol{j}})_{\alpha,\beta} c_{\boldsymbol{j},\beta}.
\end{align}
The first term describes the nearest-neighbor hopping, where $t$ is the hopping integral. 
The second term describes the chemical potential $\mu_{\ve{i}}$ at each lattice site~$\ve{i}$, and the potential barrier $V_{\ve{i}}>0$ present in the insulating MgO layers.
The third term gives rise to an attractive on-site interaction in the superconducting V layer described by the onsite potental $U_{\ve{i}}>0$.
The fourth term introduces a local magnetic exchange field $\ve{h}_{\ve{i}}$ giving rise to ferromagnetism in the Fe layer. The Pauli matrices are contained in the vector $\boldsymbol{\sigma}$.
The last term describes the Rashba spin-orbit coupling boosted by the MgO layers, where the spin-orbit field has a magnitude $\lambda_{\ve{i}}$ and is directed along the interface normal $\ve{n}$. The vector $\boldsymbol{d}_{\boldsymbol{i},\boldsymbol{j}}$ connects site $\boldsymbol{i}$ and $\boldsymbol{j}$.
In the above Hamiltonian, $c_{\boldsymbol{i},\sigma}^\dagger$ and $c_{\boldsymbol{i},\sigma}$ are the second-quantization electron creation and annihilation operators at site $\boldsymbol{i}$ with spin $\sigma$, and $n_{\boldsymbol{i},\sigma}\equiv c_{\boldsymbol{i},\sigma}^\dagger c_{\boldsymbol{i},\sigma}$ is the number operator.
The superconducting term is treated by a mean-field approach assuming that $c_{\boldsymbol{i},\uparrow} c_{\boldsymbol{i},\downarrow} \approx \left< c_{\boldsymbol{i},\uparrow} c_{\boldsymbol{i},\downarrow} \right> +\delta$, where terms to the second order in the fluctuations~$\delta$ are negligible.
The superconducting gap is defined as $\Delta_{\boldsymbol{i}}\equiv U_{\boldsymbol{i}}\left<c_{\boldsymbol{i},\uparrow}c_{\boldsymbol{i},\downarrow}\right>$ and must be treated self-consistently.
\lina{The above model is valid in the ballistic limit, but since the effects considered here depend on the formation of $s$-wave odd-frequency triplets that are robust to impurity scattering we would obtain qualitatively the same results in the diffusive limit.}

We consider a cubic lattice of size $N_x \times N_y \times N_z$ with an interface normal along the $x$ axis. We assume periodic boundary conditions in the $y$ and $z$ directions, and apply the Fourier transform
\begin{equation}
    \label{FT}
    c_{\boldsymbol{i},\sigma}=\frac{1}{\sqrt{N_y N_z}}\sum_{\boldsymbol{k}} c_{i,\boldsymbol{k},\sigma} e^{\mathrm{i}(\boldsymbol{k}\cdot\boldsymbol{i}_{||})}
\end{equation}
along these axes. To simplify notation we have defined $i\equiv i_x$, $j\equiv j_x$, $\boldsymbol{i}_{||}=(i_y , i_z )$, and $\boldsymbol{k}\equiv (k_y, k_z)$. We also use that
\begin{equation}
    \label{rel}
    \frac{1}{N_y N_z}\sum_{\boldsymbol{i}_{||}} e^{\mathrm{i}(\boldsymbol{k}-\boldsymbol{k}')\cdot\boldsymbol{i}_{||}}=\delta_{\boldsymbol{k} , \boldsymbol{k}'}.
\end{equation}
The Hamiltonian can be written on the form
\begin{equation}
    H=H_0 + \frac{1}{2}\sum_{\boldsymbol{k}}W_{\boldsymbol{k}}^\dagger H_{\boldsymbol{k}} W_{\boldsymbol{k}},
\end{equation}
where the basis is given by
\begin{align}
    W_{\boldsymbol{k}}^\dagger =& [B_{1,\boldsymbol{k}}^\dagger,...,B_{i ,\boldsymbol{k}}^\dagger ,...,B_{N_x ,\boldsymbol{k}}^\dagger ],\\\nonumber \hspace{0.5cm} B_{i , \boldsymbol{k}}^{\dagger}=&[c_{i , \boldsymbol{k} ,\uparrow}^{\dagger} \hspace{2mm} c_{i , \boldsymbol{k} ,\downarrow}^{\dagger} \hspace{2mm} c_{i , -\boldsymbol{k} ,\uparrow} \hspace{2mm} c_{i , -\boldsymbol{k} ,\downarrow}],
\end{align}
and where the Hamiltonian matrix $H_{\vek}$ consists of $N_x \times N_x$ blocks
    \begin{align}
    \label{Hamiltonian2}
        H_{i , j , \boldsymbol{k}}&= \epsilon_{i ,j ,\boldsymbol{k}} \hat{\tau}_3 \hat{\sigma}_0 +\delta_{i ,j}\Big[\mathrm{i}\Delta_{i}\hat{\tau}^+ \hat{\sigma}_y  -\mathrm{i}\Delta_{i}^* \hat{\tau}^- \hat{\sigma}_y\\\nonumber
 &+h_{i}^x \hat{\tau}_3 \hat{\sigma}_x + h_{i}^y \hat{\tau}_0 \hat{\sigma}_y +h_{i}^z \hat{\tau}_3 \hat{\sigma}_z\\\nonumber
         &-\lambda_{i}\sin(k_y )\hat{\tau}_0 \hat{\sigma}_z +\lambda_{i}\sin(k_z )\hat{\tau}_3 \hat{\sigma}_y \Big],
    \end{align}
with row and column indices $(i,j)$.
Above, $\hat{\tau}_i \hat{\sigma}_j \equiv \hat{\tau}_i \otimes\hat{\sigma}_j$ is the Kronecker product of the Pauli matrices spanning Nambu and spin space, $\hat{\tau}^\pm \equiv(\hat{\tau}_1 \pm i\hat{\tau}_2 )/2$, and
\begin{align}
    \epsilon_{i , j ,  \boldsymbol{k}} &\equiv -2t\left[\cos(k_y)+\cos(k_z)\right]\delta_{i , j}\\\nonumber
    &-t(\delta_{i , j +1}+\delta_{i , j -1}) - (\mu_{i}-V_i )\delta_{i , j}.
\end{align}
The constant term is given by
\begin{align}
    H_0&=- \sum_{i , \boldsymbol{k}}\left\{2t\left[\cos(k_y)+\cos(k_z)\right]+\mu_{i}-V_i \right\}\\\nonumber
   &+N_y N_z \sum_{i}\frac{|\Delta_{i}|^2}{U_{i}}.
\end{align}
\label{H0}
By diagonalizing $H_{\vek}$, we obtain eigenvalues $E_{n,\boldsymbol{k}}$ and eigenvectors
\begin{align}
    \Phi_{n,\boldsymbol{k}}^{\dagger}&=[ \phi_{1,n,\boldsymbol{k}}^{\dagger} \hspace{3mm} \cdots \hspace{3mm}\phi_{N_x ,n,\boldsymbol{k}}^{\dagger}],\\\nonumber
    \phi_{i ,n,\boldsymbol{k}}^{\dagger}&=[u_{i ,n,\boldsymbol{k}}^{*}\hspace{1mm} v_{i ,n,\boldsymbol{k}}^{*}\hspace{1mm} w_{i ,n,\boldsymbol{k}}^{*}\hspace{1mm} x_{i ,n,\boldsymbol{k}}^{*}].
\end{align} 
The diagonalized Hamiltonian can be written as
\begin{equation}
    H=H_0-\frac{1}{2}\sum_{n,\vek}^{'}E_{n,\vek}+\sum_{n, \boldsymbol{k}}^{'}E_{n, \boldsymbol{k}}\gamma_{n, \boldsymbol{k}}^\dagger \gamma_{n, \boldsymbol{k}},
    \label{eq:diagonal_H_independent}
\end{equation}
where the marked sum goes over 
$\{n,k_y,k_z>0\}$, $\{n,k_y>0,k_z=0,-\pi\}$, and $\{n \text{ corresponding to }E_{n,k_y ,k_z}>0,k_y=0,-\pi, k_z =0,-\pi\}$.
Expectation values of the new operators can now be evaluated according to
\begin{align}
\left<\gamma_{n,\vek}^{\dagger}\gamma_{m,\vek}\right>&=f(E_{n,\vek})\delta_{n,m},\\\nonumber
\left<\gamma_{n,\vek}^{\dagger}\gamma_{m,\vek}^{\dagger}\right>&=\left<\gamma_{n,\vek}\gamma_{m,\vek}\right>=0,
\end{align}
where $f(E_{n,\vek})$ is the Fermi-Dirac distribution.
The new quasi-particle operators are related to the old operators by
\begin{align}
    c_{i_ ,\boldsymbol{k} ,\uparrow}&=\sum_n u_{i_ ,n, \boldsymbol{k}}\gamma_{n,\boldsymbol{k}},\hspace{0.5cm}
    c_{i ,\boldsymbol{k} ,\downarrow}=\sum_n v_{i ,n, \boldsymbol{k}}\gamma_{n,\boldsymbol{k}},\\\nonumber
    c_{i ,-\boldsymbol{k} ,\uparrow}^{\dagger}&=\sum_n w_{i ,n, \boldsymbol{k}}\gamma_{n,\boldsymbol{k}},\hspace{0.5cm}
    c_{i ,-\boldsymbol{k} ,\downarrow}^{\dagger}=\sum_n x_{i ,n, \boldsymbol{k}}\gamma_{n,\boldsymbol{k}}.
\end{align}
\lina{The eigenenergies $E_{n,\vek}$ and eigenvectors $\Phi_{n,\vek}$ obtained in this diagonalization, can be used to calculate physical observables for the system.}
The superconducting gap is given by
\begin{equation}
    \Delta_i =\frac{U_i}{N}\sum^{'}_{n,\vek}\{u_{i,n,\vek}x_{i,n,\vek}^*[1-f(E_{n,\vek})]
    +v_{i,n,\vek}w_{i,n,\vek}^* f(E_{n,\vek})\},
\end{equation}
\lina{and is treated self-consistently.}

We can calculate the critical field $H_{\text{OOP}}$ for reorienting the magnetization from an IP to an OOP orientation.
The Zeeman energy of an external magnetic field $\ve{H}$ is given by
\begin{align}
    F_{\text{Zeeman}}=-\mu_0 \ve{\mu}_{\text{tot}}\cdot\ve{H},
\end{align}
where $\mu_0$ is the vacuum permeability, $\boldsymbol{\mu}_{\text{tot}}$ is the total magnetic moment, and $\ve{H}$ is the applied field.
If we consider a system where the free energy is minimal for an IP magnetization and maximal for an OOP magnetization, and we want to find the external magnetic field needed to reorient the magnetization to the OOP direction, we must require that $|F_{\text{Zeeman}}|\ge F_{\text{OOP}}-F_{\text{IP}}$. We can then calculate the critical field from
\begin{equation}
H_{\text{OOP}}=\frac{F_{\text{OOP}}-F_{\text{IP}}}{\mu_0 \mu_{\text{tot}}}.
\end{equation}
To take into account other anisotropy contributions not covered by this model, we let $F_{\text{OOP}}\to F_{\text{OOP}}+K_{\text{anis}}$. Above, the free energy is given by
\begin{equation}
    \label{F}
    F=H_0 -\frac{1}{2}\sum_{n,\vek} ^{'}E_{n,\vek}-\frac{1}{\beta}\sum^{'}_{n,\boldsymbol{k}}\ln(1+e^{-\beta E_{n, \boldsymbol{k}}}),
\end{equation}
where $\beta=(k_B T)^{-1}$.
The total magnetic moment of the system for an OOP magnetization is given by
\begin{align}
    \mu_{\text{tot}} &=  -2\mu_{\text{B}}\sum_{i,n,\vek}^{'}\big\{\Real(u_{i,n,\vek}^* v_{i,n,\vek})f(E_{n,\vek})\\\nonumber
    &+\Real(x_{i,n,\vek}^* w_{i,n,\vek})\left[1-f(E_{n,\vek})\right]\big\},
\end{align}
when the interface normal is directed along the $x$ axis.

Since the lattice must be scaled down in order to make the system computationally manageable, we choose the magnitude of the on-site coupling potential $U_{i}$ so that the superconducting coherence length is comparable to the thickness of the V layer. The superconducting coherence length is given by $\xi=\hbar v_F /\pi\Delta_0$, where $v_F=(1/\hbar)dE_{\boldsymbol{k}}/dk\big|_{k=k_F}$ is the Fermi velocity calculated for the normal-state eigenenergy $E_{\boldsymbol{k}}=-2t[\cos(k_x )+\cos(k_y )+\cos(k_z )]-\mu$, and $\Delta_0$ is the zero-temperature superconducting gap.

We determine the superconducting critical temperature by a binomial search, where we decide if a given temperature is above or below $T_C$. This is decided by finding whether the superconducting gap measured in the middle of the superconducting region increases towards a superconducting solution or decreases towards a normal state solution from the initial guess $\Delta\ll\Delta(T=0)$ under iterative recalculations. 

In Fig. \ref{fig:BdG_theory}, we have used the parameters $t=1$, $\mu_{i\in\text{S}}=0.9$, $\mu_{i\in\text{SOC}}=\mu_{i\in\text{F}}=0.8$, $V_{i\in\text{SOC}}=0.79$, $U=1.4$, $\lambda=0.4$, $h=0.8$, $N_x^{\text{S}}=28$, $N_x^{\text{SOC}}=3$, $N_x^{\text{F}}=8$, and $N_y = N_z = 50$. This gives a coherence length of $21$ lattice sites. All length scales are scaled by the lattice constant $a$, all energy scales are scaled by the hopping parameter $t$, and the magnitude of the spin-orbit coupling is scaled by $ta$.

\end{document}